\documentclass{svjour2}

\usepackage{graphicx}

\usepackage{rotating}
\usepackage{amssymb}
\usepackage{mathptmx}
\usepackage{upgreek}
\usepackage[numbers]{natbib}

\begin{document}

\title{Quartz Tuning Forks and Acoustic Phenomena -- Application to Superfluid Helium}

\author{J. Rysti \and J. Tuoriniemi}

\institute{O.V. Lounasmaa (Low Temperature) Laboratory, Aalto University, P.O. Box 15100, 00076 Aalto, Finland\\
\email{juho.rysti@aalto.fi}
}

\date{}

\maketitle

\begin{abstract}
Immersed mechanical resonators are well suited for probing the properties of fluids, since the surrounding environment influences the resonant characteristics of such oscillators in several ways. Quartz tuning forks have gained much popularity in recent years as the resonators of choice for studies of liquid helium. They have many superior properties when compared to other oscillating bodies conventionally used for this purpose, such as vibrating wires. However, the intricate geometry of a tuning fork represents a challenge for analyzing their behavior in a fluid environment -- analytical approaches do not carry very far. In this article the characteristics of immersed quartz tuning fork resonators are studied by numerical simulations. We account for the compressibility of the medium, that is acoustic phenomena, and neglect viscosity, with the aim to realistically model the oscillator response in superfluid helium. The significance of different tuning fork shapes is studied. Acoustic emission in infinite medium and acoustic resonances in confined volumes are investigated. The results can aid in choosing a quartz tuning fork with suitable properties for experiments, as well as interpreting measured data.

\keywords{Compressible fluid \and Acoustic emission \and Acoustic resonance}
 \PACS{47.80-v \and 67.90.+z \and 85.50.-n}
\end{abstract}

\section{Introduction}

Quartz tuning fork resonators have become popular tools for studying superfluid helium. They have been used to detect, for example, temperature, pressure, density, concentration, viscosity, and turbulence \cite{springerlink:10.1023/B:JOLT.0000035368.63197.16, springerlink:10.1007/s10909-006-9279-4, springerlink:10.1007/s10909-007-9389-7, springerlink:10.1007/s10909-007-9587-3, springerlink:10.1007/s10909-011-0394-5, Bradley2009, Bradley2013}. To a great extent, they have replaced other mechanical resonators, such as vibrating wires. Their main advantages are availability through mass production, ease of use, supreme stability, and high intrinsic quality factor. The major disadvantage is their tricky geometry, which restricts analytical studies of the resonant behavior in fluid environments. All quartz tuning forks have the same two-pronged shape, but they come in variety of sizes, relative dimensions, and resonant frequencies. The most common room temperature resonant frequency of a quartz tuning fork in vacuum is designed to be $2^{15} = 32768$~Hz, but availability of forks with other frequencies allow experiments to be done in a wide range of acoustic wavelengths. The forks are driven by the piezoelectric effect on quartz, which also creates measurable electric currents once the quartz prongs bend periodically at the mechanical resonance. Mass produced quartz tuning forks are encapsulated in hermetically sealed cylindrical metal cans or parallelepiped caskets. To make contact between the resonator and fluid, either holes must be created in the capsule or the container be completely removed.

The resonant frequency $f_0$ of a mechanical oscillator immersed in inviscid fluid medium is given by \cite{springerlink:10.1007/s10909-006-9279-4}
\begin{equation} \label{Eq:Bdefinition}
f_0 = f_{\mathrm{0vac}} \left(1 + B \frac{\rho_{\mathrm{F}}}{\rho_{\mathrm{S}}} \right)^{-1/2} ,
\end{equation}
where $f_{\mathrm{0vac}}$ is the resonant frequency in vacuum, $B$ is a geometry-dependent dimensionless factor of order unity, and $\rho_{\mathrm{F}}$ and $\rho_\mathrm{S}$ are the densities of the fluid and the oscillator (solid), respectively. The density of quartz is $\rho_\mathrm{S} = 2659$~kg/m$^3$. For compressible media, the factor $B$ depends on the wavelength of sound in the fluid. If the wavelength is much larger than the relevant oscillator dimensions, the medium is often assumed to be incompressible. Under this approximation, $B$ is a constant, which depends only on the geometry of the system. In precise measurements this assumption cannot be made even if the wavelength is an order of magnitude larger than the dimensions. The situation becomes even worse if acoustic resonances in the cavity around the oscillator, or within the oscillator structures, are excited. In the vicinity of an acoustic resonance, $B$ changes rapidly as a function of wavelength. The geometrical factor can be calculated analytically only for some simple geometries, such as an infinitely long circular or elliptic cylinder. It is one goal of this paper to determine, how $B$ depends on various tuning fork properties and configurations.

We examine the resonant behavior of quartz tuning forks by numerical simulations using the finite element method (FEM). We study the properties of tuning forks with various prong shapes and compute the resonant frequency change due to the surrounding compressible fluid. We consider the oscillators in infinite medium, as well as in a confined cylindrical cavity. In infinite medium we compute the emission of acoustic radiation and in confined space we consider the acoustic resonant modes within the fluid volume. The influence of acoustic impedance on the surfaces is also discussed. The significance of different cross-sectional shapes is studied by assuming the oscillator to be infinitely long, which reduces the calculation to two dimensions. This is computationally much less demanding than the full 3D models. The full fork geometry is simulated to make comparison with experiments and to examine the difference between the 2D and 3D models. We have previously reported the results of similar studies for various 2D geometries \cite{Compressibility1}.

\section{Computational Model}

In our simulations we have mainly used an effective method, whose validity was checked against a more comprehensive model. The effective method is computationally much less demanding and can produce results considerably faster. In this chapter we describe the simulation procedure. We approximate the tuning fork as a harmonic oscillator and therefore first quickly review the relevant analysis of a general harmonic system before turning to the tuning fork.

\subsection{Harmonic Oscillator}

For small oscillations of a resonator, we can use the equation of motion of a driven damped harmonic oscillator
\begin{equation} \label{Eq:harmonic}
F_\mathrm{e} = m \frac{d^2 x}{d t^2} + m \gamma \frac{dx}{dt} + k_\mathrm{S} x ,
\end{equation}
where $F_\mathrm{e}$ is the excitation force, $m$ is the oscillator mass, $x$ is the displacement, $k_\mathrm{S}$ is the spring constant, and $\gamma$ represents damping. If the oscillator is in vacuum, $\gamma$ stands for intrinsic damping of the oscillator. A finite value for $\gamma$ is applied for computational convenience. We seek time-harmonic solutions of Eq.~(\ref{Eq:harmonic}), where all the variables are proportional to $e^{i \omega t}$, that is, they oscillate sinusoidally at a frequency $\omega$. This eliminates time from the equations. We denote $x = X e^{i \omega t}$, $F_\mathrm{e} = F_\mathrm{E} e^{i \omega t}$, and so on. Separating the real and imaginary components of the amplitude, $X = a + b i$, gives
\begin{equation} \label{Eq:real_part}
a = - \frac{F_\mathrm{E}}{m} \frac{\omega^2 - \omega_0^2}{(\gamma \omega)^2+(\omega^2 - \omega_0^2)^2}
\end{equation}
and
\begin{equation}
b = - \frac{F_\mathrm{E}}{m} \frac{\gamma \omega}{(\gamma \omega)^2+(\omega^2 - \omega_0^2)^2} .
\end{equation}
Here $\omega_0^2 \equiv k_\mathrm{S}/m$ is the resonant frequency squared. We can represent these equations in a form appropriate for the "single frequency measurement method", also used experimentally \cite{springerlink:10.1007/s10909-011-0394-5}, by solving the width $\Delta f = \gamma/2 \pi$ and resonant frequency $f_0 = \omega _0/2 \pi$ as functions of frequency $f = \omega/2\pi$ and the real and imaginary parts of the amplitude:
\begin{equation} \label{Eq:gamma}
\Delta f = - \frac{F_\mathrm{E}}{m} \frac{1}{4 \pi^2f} \frac{b}{a^2 + b^2}
\end{equation}
\begin{equation} \label{Eq:omega}
f_0 =f \sqrt{1 - \frac{a}{b}\frac{\Delta f}{f}}
\end{equation}
We can now vary some properties of the system, for example the speed of sound in the fluid, and simply use an arbitrary "measurement frequency" $f$ to determine the proper resonant frequency. Non-linear frequency dependence of the oscillator response on the fluid motion causes a slight computational inconvenience. If one uses Eqs. (\ref{Eq:gamma}) and (\ref{Eq:omega}), the "measurement frequency" should be quite close to the resonance. This is ensured by iterating $f$ a few times and having it depend on the system property being varied, in our case the speed of sound. The necessity of having $f$ close to $f_0$ is particularly profound if the oscillator is coupled to an acoustic resonance of the fluid.

\subsection{Fluid Environment}

We model the surrounding fluid as an acoustic medium, which obeys the Helmholtz equation
\begin{equation} \label{Eq:helmholtz}
\nabla^2 P - k^2 P = 0 ,
\end{equation}
where $P$ is the acoustic pressure (deviation from some equilibrium value) and $k = \omega/c = 2 \pi / \lambda$ is the acoustic wave vector corresponding to wavelength $\lambda$. Incompressible fluid is obtained in the limit $k \rightarrow 0$ ($\lambda \rightarrow \infty$). Speed of ordinary sound (first sound) in helium liquids vary between 100~m/s and 400~m/s, depending on temperature, pressure, and concentration in mixtures. A 32.8~kHz quartz tuning fork can therefore cover a wave vector range between $k=500$~m$^{-1}$ and $k=2000$~m$^{-1}$. Gaseous helium extends the range to much higher values. Higher-frequency oscillators allow experiments to be performed at wave vectors in the several thousands~m$^{-1}$.

The oscillating object directs a normal acceleration on the fluid in the direction of its motion. The boundary condition is thus
\begin{equation}
\hat{n} \cdot \nabla P = \rho_\mathrm{F} \ddot{X} ,
\end{equation}
where $\hat{n}$ is the unit normal vector of the boundary and $\ddot{X}$ is the "acceleration amplitude" defined through $d^2x/dt^2 = \ddot{X} e^{i \omega t}$. The reaction force $F_\mathrm{M}$ of the medium on the oscillator is obtained by integrating the pressure field of the fluid over the oscillator surface. The acceleration of the resonator $\ddot{X}$ is solved from the harmonic oscillator equation, Eq.~(\ref{Eq:harmonic}), by including $F_\mathrm{M}$ in the equation of motion:
\begin{equation}
F_\mathrm{E} = \left(m - i \frac{\Delta f_{\mathrm{vac}}}{f} m - \frac{k_\mathrm{S}}{(2 \pi f)^2} \right) \ddot{X} - F_\mathrm{M} .
\end{equation}
One should note that now $\Delta f$, which is solved from Eq.~(\ref{Eq:gamma}), can be different from $\Delta f_{\mathrm{vac}}$, which is given in as a parameter describing intrinsic losses in the oscillator. The displacement is simply given by
\begin{equation} \label{Eq:displacement}
X = \frac{ \ddot{X}}{(2 \pi f)^2} .
\end{equation}
The boundary condition on the container walls and non-accelerating surfaces of the oscillator is
\begin{equation}
\hat{n} \cdot \nabla P = -P \rho_\mathrm{F} \frac{i \omega}{Z} ,
\end{equation}
where $Z$ is the surface acoustic impedance. The limit $Z \rightarrow \infty$ corresponds to the perfectly reflecting hard boundary. The surface impedance can be written as \cite{RoomAcoustics}
\begin{equation}
Z = \rho_\mathrm{F} c_\mathrm{F} \frac{1+R}{1-R} ,
\end{equation}
where $R$ is the amplitude reflection coefficient. For a completely absorbent interface, $R=0$, the surface impedance equals the characteristic impedance of the fluid medium $\rho_\mathrm{F} c_\mathrm{F}$. In the case of an interface between two bulk media, $R$ is given as \cite{LandauLifshitz}
\begin{equation}
R = \frac{\rho_\mathrm{S} c_\mathrm{S} - \rho_\mathrm{F} c_\mathrm{F}}{\rho_\mathrm{S} c_\mathrm{S} +\rho_\mathrm{F} c_\mathrm{F}} .
\end{equation}
Infinite medium is simulated by applying a second order expansion cylindrical wave radiation boundary condition (spherical wave in 3D) at an artificial boundary far from the oscillator \cite{CylindricalSphericalWave}. This type of condition tends to minimize reflections at the boundaries, thus mimicking infinite medium.

\subsection{Quartz Tuning Fork} \label{Sec:Fork}

\begin{figure}
\begin{center}
\includegraphics[%
  width=0.9\linewidth,
  keepaspectratio]{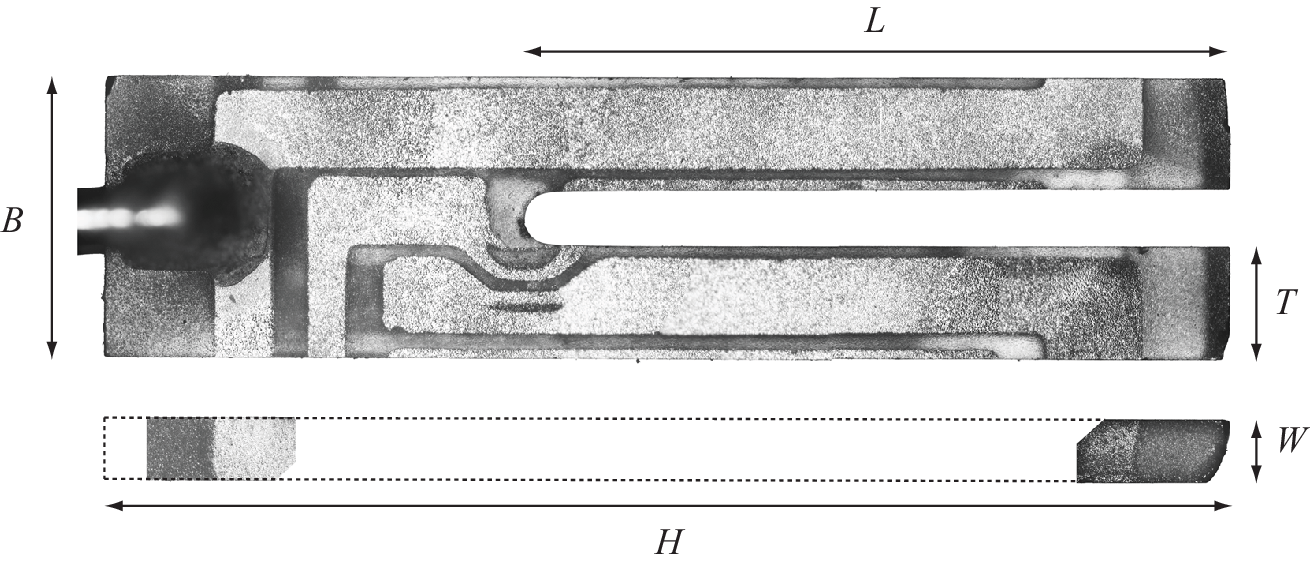}
\end{center}
\caption{Microscope image of a Fox Electronics NC38 quartz tuning fork, similar to the one used in the experiments of references \cite{springerlink:10.1007/s10909-007-9583-7} and \cite{PhysRevB.78.064509}. The lower image shows the side-view of the tuning fork, but the oscillator has not been photographed along its entire length from this direction. Length of the tines is $L=3.76$~mm, width of the tuning fork $W=0.34$~mm, tines in the other direction $T=0.60$~mm, base $B=1.51$~mm, and the total height $H=6.00$~mm. The electrodes used for excitation and detection of the resonance are visible on the surface of quartz. Electrical contact is achieved with wires soldered to the electrodes on both sides of the base.}
\label{Fig:NC38}
\end{figure}

Experimental data on the geometrical factor $B$ for a quartz tuning fork in helium fluids exist \cite{springerlink:10.1007/s10909-007-9583-7,PhysRevB.78.064509}. The oscillator used in those experiments was a Fox Electronics NC38 quartz tuning fork. Unfortunately the particular fork used in the experiments was not available for closer inspection, so the exact dimensions and configuration of the device could not be precisely determined. We did obtain, however, other items from the same manufacturing batch to measure their properties. One NC38 quartz tuning fork was removed completely from its casing in order to examine the geometry of the oscillator. A microscope image of that tuning fork and the measured dimensions are shown in Fig.~\ref{Fig:NC38}.

The inner dimensions of the cylindrical container were inspected by puncturing a small hole into the container of another NC38 and filling the volume with epoxy, after which the metallic can was corroded away by acid. It was then possible to use a microscope to determine the dimensions. The cavity diameter was found to be 2.63~mm and the distance between the tine ends and the container "ceiling" 1.12~mm. The total cavity height used in the simulations was 7.12~mm. The space between the tuning fork base and the container "floor" could not be determined very accurately, but it was of the order of 0.1~mm. That small space was neglected in the simulations, as its effect is minuscule and meshing such a small volume with FEM elements would require a rather fine mesh and thus increase the number of degrees of freedom to little avail.

By filling the cavity with epoxy, we could also see how the tuning fork was positioned inside the container. The epoxy-filled tuning fork had a clearly observable tilt of about 5 degrees in the direction perpendicular to the oscillation. Any possible tilt in the other direction was hard to observe, as the epoxy cast caused too much distortion. In its designed operation, the quartz tuning fork is in vacuum and thus its orientation inside the casing does not matter. When the container is breached and fluid fills the cavity, the exact orientation makes considerable difference, as will be demonstrated in Sec.~\ref{Sec:AcousticResonances}.

The tines of a quartz tuning fork are effectively cantilever beams, and they must be analyzed accordingly. The tines can oscillate in a great variety of different resonant modes with increasing frequency. The lowest-frequency mode, which can be excited by the electrodes on the surface of the fork, is such that the tines oscillate in antiphase toward each other and no nodes exist along the tines. This is also the typical mode used in most applications. Experiments with higher-frequency modes of quartz tuning forks have also been performed, but here we consider only the lowest usable mode. The fork can be described with an effective harmonic oscillator model using Eq.~(\ref{Eq:harmonic}) and replacing the mass of the oscillator (one tine) with an effective mass given by \cite{ForkEffective}
\begin{equation}
m^* = 0.2427 m .
\end{equation}
This applies to a uniform rectangular cantilever beam. If the shape or density distribution of the tine is something else, the effective mass must be recalculated. The spring constant in the harmonic oscillator equation is given by
\begin{equation}
k_\mathrm{S} = \frac{E}{4} W \left( \frac{T}{L} \right)^3 ,
\end{equation}
where the dimensions are as in Fig.~\ref{Fig:NC38} and $E$ is the elastic modulus. For quartz we have $E=7.87 \cdot 10^{10}$~Pa. These give a vacuum resonant frequency for the NC38 as $f_{\mathrm{0vac}} = \sqrt{k_\mathrm{S}/m^*}/2\pi = 37 300$~Hz. This is 14~\% larger than the measured value (32770~Hz) at room temperature. Such discrepancy has been attributed to additional mass of the electrodes, difference in the elastic modulus, and deviation from the ideal geometry \cite{springerlink:10.1007/s10909-006-9279-4}. It seems, however, that the discrepancy is in fact mostly due to the simplifying assumptions of the basic cantilever model. Fixing the beams right from one end raises the resonant frequency compared to a full quartz tuning fork model, where the oscillator beams are fixed at the common base and the base also flexes a little. We determined the eigenfrequencies of the complete tuning fork geometry by using a linear elastic material model and Comsol Multiphysics software. So obtained resonant frequency is 30790~Hz, which is 6~\% smaller than the measured value and 18~\% smaller than the one given by the basic cantilever model. By fixing the tines of the full tuning fork model similarly as in the basic model at the base connection, the resonant frequency of the simple model was reproduced. If we take into account the actual circular base shape between the tines, the obtained frequency is 31770~Hz, which brings the difference between the measured and calculated values to 3~\%. The remaining difference can be attributed with confidence to the other effects mentioned before.

In the effective model the excitation force is applied to the tips of the tines. Correspondingly, the acoustic force due to pressure of the surrounding fluid must be properly scaled to be compatible with this. The appropriate scaling factor is obtained by weighing the position-dependent force by the displacement $x$ of the cantilever beam at height $z$ from the tine's fixing point \cite{SecondSoundModes}
\begin{equation} \label{Eq:EulerBeam}
x/x_0 = \frac{1}{3} \left(z/z_0 \right)^4 - \frac{4}{3} \left(z/z_0 \right)^3 + 2 \left(z/z_0 \right)^2 .
\end{equation}
Here $z$ assumes values between $0$ and $z_0$, and $x$ correspondingly between $0$ and $x_0$. The normal acceleration applied to the fluid must also be scaled by this along the fork tines, as can be deduced from Eq.~(\ref{Eq:displacement}).

The constructed effective model of the quartz tuning fork was compared to a "full model", where we meshed the tuning fork as well and treated the system as an acoustic-solid interaction problem in the frequency domain. The solid was modeled as linear elastic material. All damping effects were omitted ($\gamma = 0$, closed volume, no viscosity, and perfectly reflecting walls) so that we can use just the real part of the oscillation amplitude, Eq.~(\ref{Eq:real_part}), to determine the resonant frequency by calculating the response at a few frequencies. That is, we fit the oscillation amplitudes to
\begin{equation}
a = -\frac{C}{f^2 - f_0^2},
\end{equation}
where $C$ is a constant ($b=0$ except at $f=f_0$). After obtaining this constant, we can use a single computing frequency to determine the actual resonant frequency. The calculated geometrical factors $B$ of the effective and the full models agreed within $5~\%$. The difference is likely due to the fixed-cantilever-beam approximation mentioned above. This will be discussed further below.

The numerical solutions were found using the finite element method, which is practical for systems with complicated geometries. The computations were performed with Comsol Multiphysics software. We used a desktop computer with 8 Gb of random access memory, which posed restrictions to the maximum number of mesh elements, ample especially in the 3D models. In particular, the "full model" and the 3D "infinite medium" proved to be on the limits for this hardware. Even with this restriction, the results obtained are considered to be quite accurate. Symmetries were utilized throughout when possible to decrease the size/number of elements and to speed up the simulations. The maximum number of elements in the 3D models was about 300 000.

\section{Results}

\subsection{Geometrical Factor}

\begin{figure}
\begin{center}
\includegraphics[%
  width=0.9\linewidth,
  keepaspectratio]{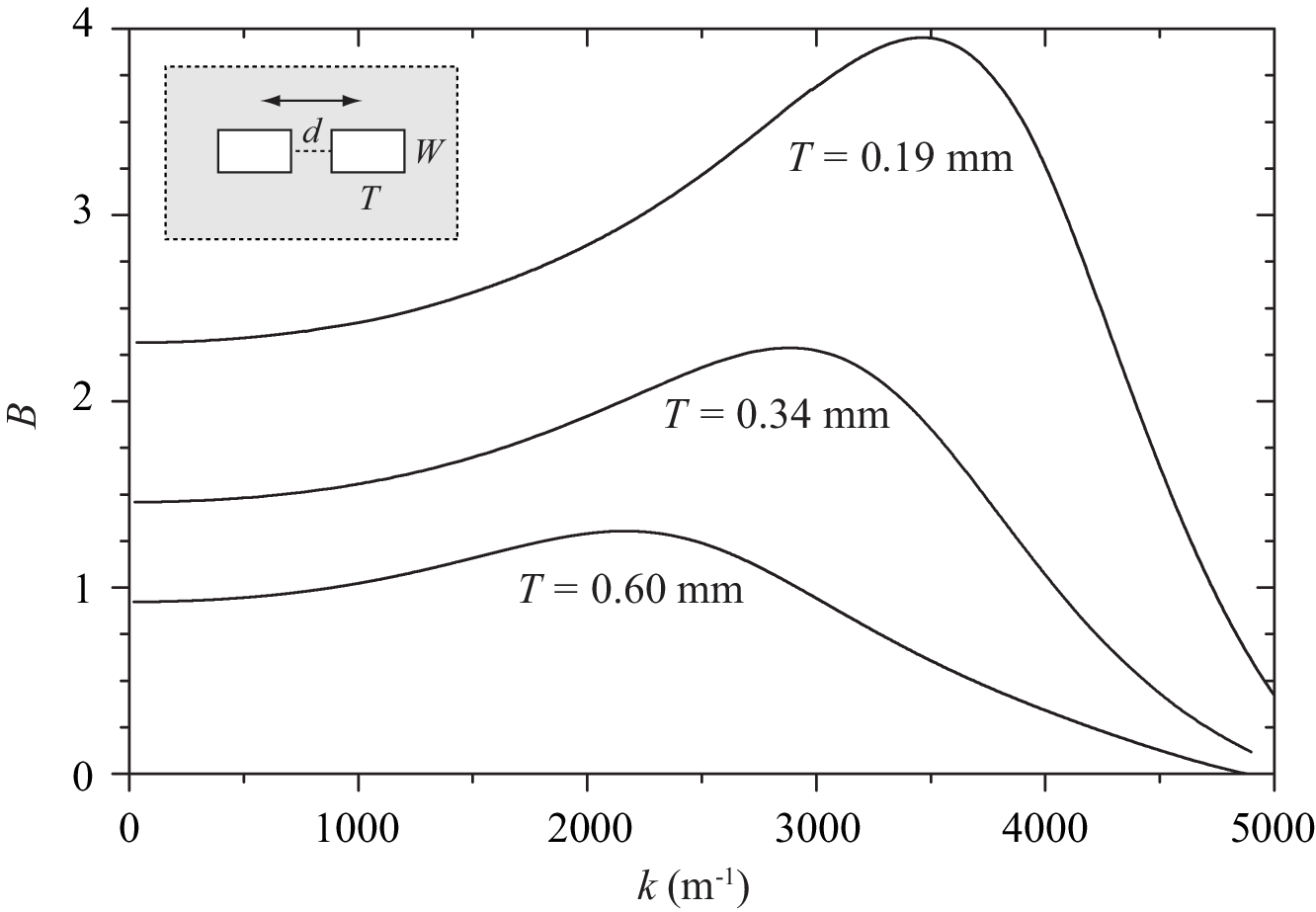}
\end{center}
\caption{Geometrical factors $B$ for a pair of infinitely long rectangular bars in infinite medium oscillating in antiphase. The geometry is depicted in the upper-left corner. The arrows represent the direction of tine motion. The relative dimensions of the oscillators are varied by changing $T$ as the separation $d = 0.30$~mm and the width $W = 0.34$~mm are kept constant. The case with $T = 0.60$~mm is the 2D version of the NC38 quartz tuning fork shown in Fig.~\ref{Fig:NC38}.}
\label{Fig:BvsShape2D}
\end{figure}

The geometrical factor $B$ is obtained from the simulated frequency data through Eq.~(\ref{Eq:Bdefinition}). We begin by reporting the results for oscillators of infinite length. This reduces the problem to two dimensions and is computationally much easier. We study the effects of different prong shapes and separations on the geometrical factor. Fig.~\ref{Fig:BvsShape2D} shows $B$ for three different relative cross-sectional dimensions of infinitely long prongs in infinite medium as functions of wave vector. If the overall size of the oscillating pair is changed, this merely scales the $k$-axis without altering the functional form of $B$.
At lower values of $k$, $B$ does not change much, but beyond $k \sim 1000$~m$^{-1}$, there is a notable increase. It reaches a maximum, whereafter the $B$-factor decreases rapidly. The wave vector of the maximum and its peak value clearly depend on the geometry. Larger $T/W$ ratio results in less pronounced variations of $B$ over the entire wave vector range. Similar studies when changing, for example, the separation $d$ between the oscillators have been reported in Ref.~\cite{Compressibility1} and are not repeated here in detail. For comparison, we note that decreasing $d$ has a similar effect to $B$ as decreasing $T$.

\begin{figure}
\begin{center}
\includegraphics[%
  width=0.9\linewidth,
  keepaspectratio]{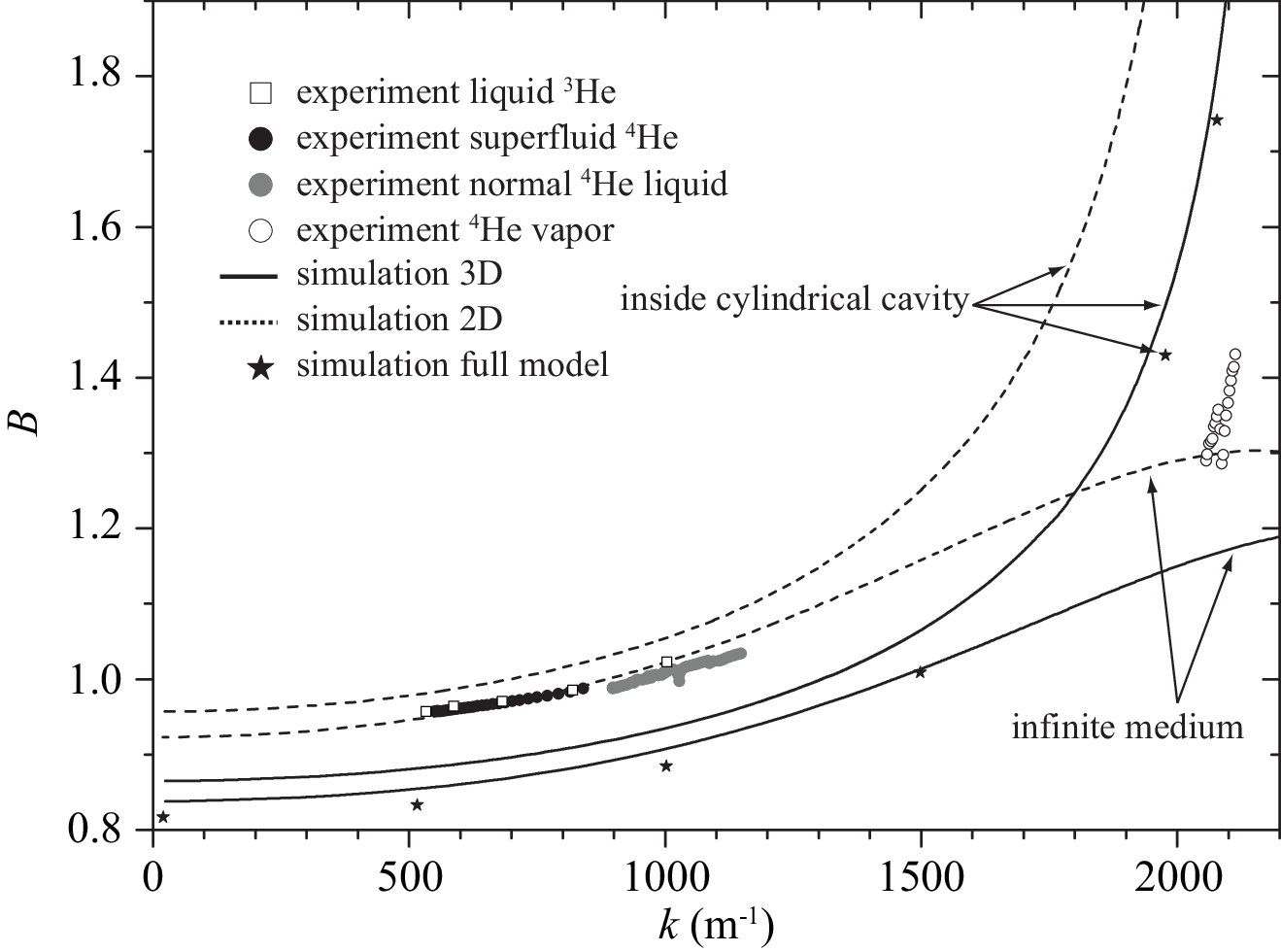}
\end{center}
\caption{Comparison between simulation and experimental data of the geometrical factor $B$. The experimental data represents results from helium fluids at four different conditions \cite{springerlink:10.1007/s10909-007-9583-7,PhysRevB.78.064509}. The data have been extrapolated to zero viscosity. The solid lines in the graph represent 3D simulations of the tuning fork geometry, using the effective model. The fork is either located symmetrically inside a cylindrical cavity or placed in infinite medium. The dashed lines are the same for 2D oscillators. The star symbols represent simulations using the "full model" of the tuning fork geometry. Dimensions of the oscillator are as in Fig.~\ref{Fig:NC38}. The diameter of the cylindrical cavity is 2.63~mm.}
\label{Fig:BCompareExptCalc}
\end{figure}

Fig.~\ref{Fig:BCompareExptCalc} shows the simulated results for $B$ of the 3D tuning fork in a cylindrical cavity and in infinite medium, together with the experimental data of Refs~\cite{springerlink:10.1007/s10909-007-9583-7} and~\cite{PhysRevB.78.064509}.
The tuning fork in the computational model is positioned symmetrically inside the cavity. For comparison, the results for 2D simulations and for the "full model" are also displayed. The overall computed values of the geometrical factor and the general dependence of $B$ on the wave vector are more or less consistent with the experiment. At lower wave vectors, $B$ increases at a slow rate and then begins to ramp up as a function of $k$. In infinite medium, the increase of $B$ is significantly less pronounced. As noted in Ref.~\cite{Compressibility1}, the steep increase of $B$ within the finite container compared to infinite medium is due to the impending acoustic resonance in the fluid-filled cavity.

The simulated values of $B$ at low wave vectors are about 7~\% smaller than those measured and revert to about 45~\% greater at large $k$. When the tuning fork is tilted relative to the container, it can excite additional acoustic resonant modes within the cavity. These modes affect $B$ and much of the discrepancy between the simulation and experiment can be explained by acoustic resonances in the container due to asymmetric positioning of the tuning fork. This will be discussed further in Sec.~\ref{Sec:AcousticResonances}. We now consider other effects, which influence $B$ even in the symmetrical situation.

The apparent imperfections in the tuning fork visible in Fig.~\ref{Fig:NC38} do not explain the difference between the measurement and simulation. Even the somewhat disfigured tips of the tuning fork, where one would expect the strongest effects, result in surprisingly small change in the geometrical factor. Some other possible effects, which would change $B$, can also be thought of, but none of them are realistically large enough to explain the observed discrepancy. Acoustic impedance of typical bulk metal, $Z \approx 3 \cdot 10^7$~Pa$\cdot$s/m, is so large that the surface is practically fully reflecting. There is no observable effect due to the impedance unless the system is very close to acoustic resonance of the cavity and the impedance is significantly smaller. If $Z$ is reduced two orders of magnitude, $B$ changes from 1.71 to 1.66 at $k = 2055$~m$^{-1}$. In the experiments, the top of the can was partially open to allow fluid to enter the cavity. Some acoustic radiation thus unavoidably escaped the container through these holes. This leads to small decreases in $B$. If the top is completely open, then $B(2055~\mathrm{m^{-1}}) = 1.68$. Only by removing large parts of the container walls and exposing the tuning fork, would there be a significant decrease in $B$ at large values of $k$, although still not large enough to reproduce the experimental values. Shrinking the radius of the container by a factor 0.98 changed $B(2055$~m$^{-1})$ to 1.64. One may also speculate about the effect of the electrodes on the surface of the quartz. Density and elastic modulus of the tines are not uniform because of that. Density variations change the effective mass, but this should be unconvincingly different to change the results in any significant way. This effect roughly just scales $B$ over the entire $k$-range, at least if the mass differences are modest. Smaller effective mass yields larger $B$.

The results obtained by the "full calculation" are persistently approximately 5~\% smaller than the ones obtained from the effective model at all values of $k$. The absolute difference in $B$ grows from 0.055 at small wave vectors to 0.067 at about $k \approx 2100$~m$^{-1}$, but the relative difference decreases at the same time from approximately 5.5~\% to 3.8~\%. The number of elements could have been inadequate in the "full model", as our computational power did not permit a finer mesh, but the disagreement did not increase significantly by making the mesh even sparser. As was discussed earlier, the discrepancy is more likely due to the effective model assuming the tines being fixed at the base end. The full model disagrees with the experiment more than the effective model at low $k$. This conclusion might be a bit hasty, however, since tilting of the tuning fork inside the capsule tends to correct the too-low values at low-$k$ and the too-high values at large-$k$, as is demonstrated in the final subsection of this chapter.

\subsection{Acoustic Emission}

If the oscillator and fluid are placed inside a closed container with perfectly (acoustically) reflecting walls, the resonant frequency changes according to Eq.~(\ref{Eq:Bdefinition}), but the width of the resonance $\Delta f$ remains the same as in vacuum. However, if dissipation of acoustic energy takes place in the system, the width increases compared to the vacuum value. Dissipation can exist, for example, at absorbing (or transmitting) boundaries or if boundaries do not exist at all (acoustic energy pumped from the oscillator into the medium does not completely return to the oscillator). We can study the acoustic emission from the fork in infinite medium by determining $\Delta f$. The excess width is directly related to dissipation in the system -- in this case to the emitted acoustic power.

\begin{figure}
\begin{center}
\includegraphics[%
  width=0.9\linewidth,
  keepaspectratio]{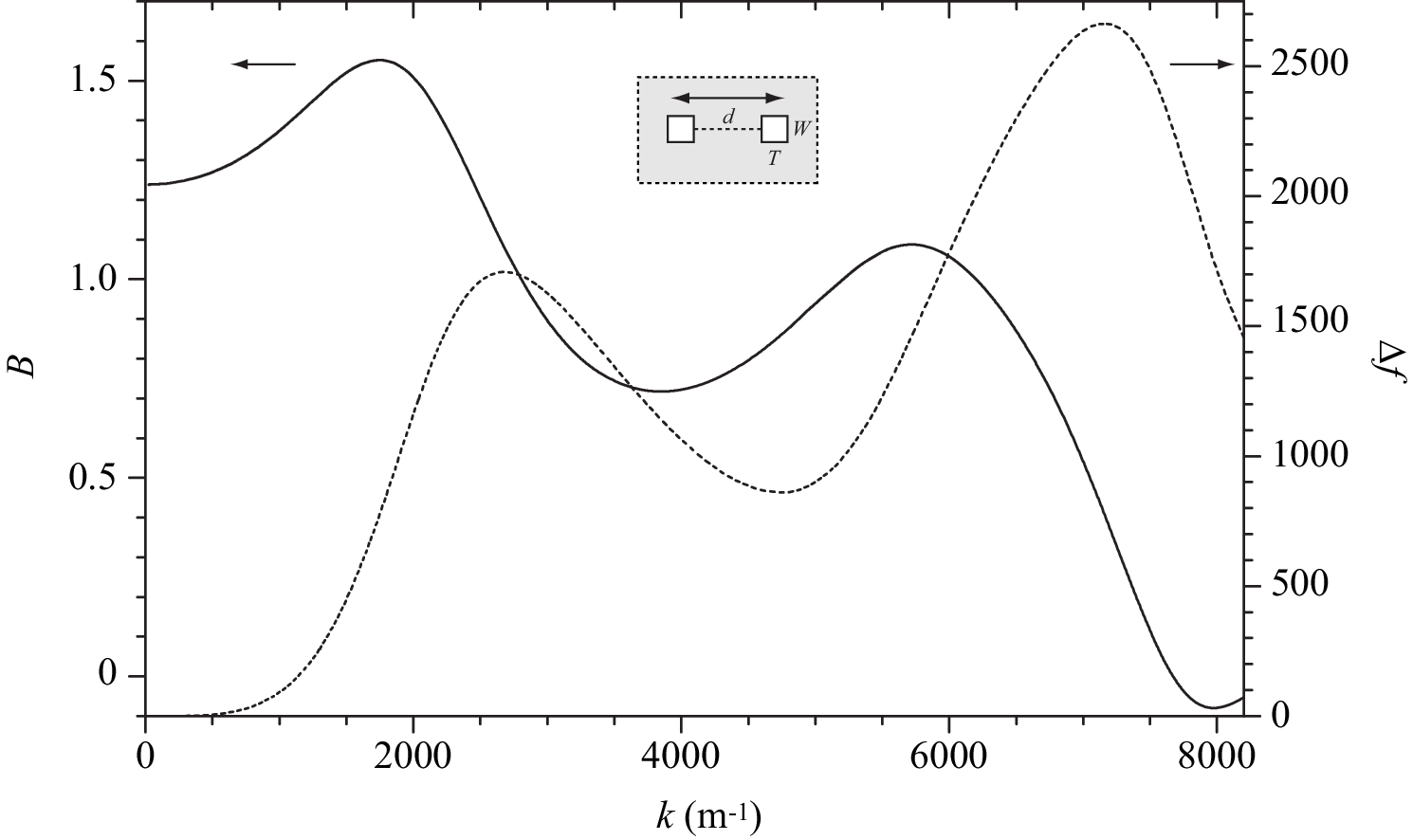}
\end{center}
\caption{The geometrical factor $B$ and the width of the resonance $\Delta f$ as functions of $k$ for two infinitely long square-shaped bars ($T=W=0.34$~mm) oscillating in antiphase in infinite medium. The separation between the oscillating bars is $d=1.0125$~mm. The solid curve is for $B$ (left vertical axis) and the dashed for $\Delta f$ (right vertical axis). The vacuum width used in the calculations (20~Hz) has been subtracted from $\Delta f$. Geometry of the system is depicted in the figure.}
\label{Fig:BAcousticTwoSquares}
\end{figure}

An example of the relation between the geometrical factor $B$ and the width $\Delta f$ is shown in Fig.~\ref{Fig:BAcousticTwoSquares}.
The figure shows the case of two infinitely long square-shaped bars separated by 1.0125~mm in infinite medium oscillating in antiphase. Rather large separation between the oscillators has been used here to make acoustic resonances appear at lower values of $k$. The undulation of $B$ and $\Delta f$ is partly due to resonant modes in the fluid between the bars. The first peak is a general property of all oscillators, and obviously reflects a matching wavelength relative to the oscillator dimensions. The subsequent peaks appear only when acoustic modes between two-part oscillators exist. Fig.~\ref{Fig:BAcousticTwoSquares} suggests an apparent connection between the changes in $B$ and $\Delta f$: the width is at maximum or minimum when the frequency changes the most and vice versa. This is not coincidental. We can examine the relationship between the two quantities by the Kramers-Kronig relations. By introducing a cutoff frequency, the wave vector values of the minima and maxima can be reproduced by calculation through the Kramers-Kronig. Precise analysis is difficult due to limited range of simulated data on such acoustic resonances.

\begin{figure}
\begin{center}
\includegraphics[%
  width=0.85\linewidth,
  keepaspectratio]{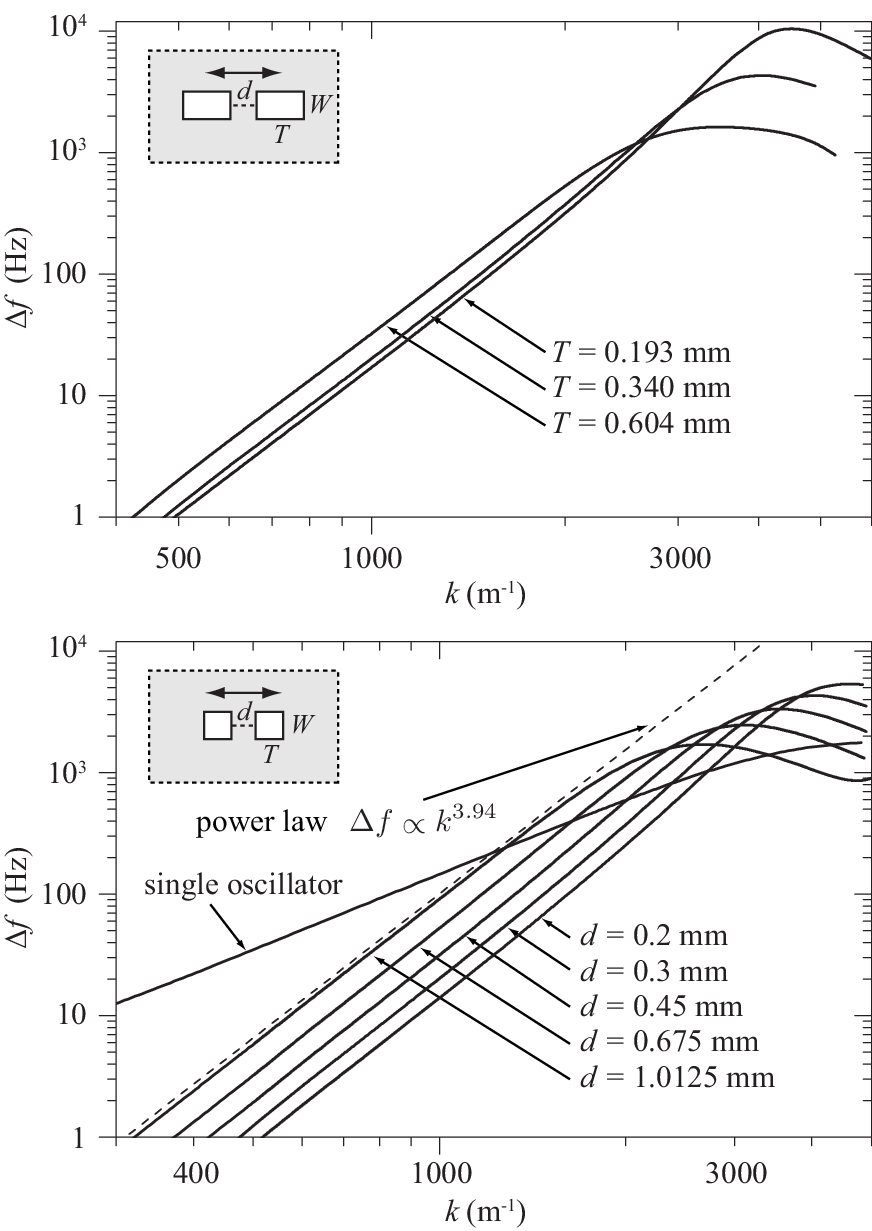}
\end{center}
\caption{Width of the resonance of infinitely long oscillators in infinite medium. The top frame shows the effect of variation of $T$ holding $d = 0.30$~mm and $W=0.34$~mm constant. The bottom frame shows the effect of variation of $d$ with fixed $T = W = 0.34$~mm. A power law fit to the low-$k$ data is plotted on the bottom frame (dashed line). A small offset (prefactor adjustment) has been applied to show the fit more clearly. The computed width for a single-rod square-shaped oscillator is displayed for reference. The other geometries are depicted in the figure. The vacuum width used in the calculations (20~Hz) has been subtracted from $\Delta f$.}
\label{Fig:AcousticPower2D}
\end{figure}

Fig.~\ref{Fig:AcousticPower2D} illustrates how the width of the resonance depends on the oscillator geometry in 2D.
On the left frame, the relative dimensions are varied by changing $T$ and keeping the other dimensions constant. On the right, the separation between the oscillating rods is varied. For comparison, the case of a single oscillating rod is also shown on the right frame. A power law $\Delta f \propto k^p$ fits the simulated data well at small wave vectors with an exponent $p=3.94$ for the double-rod geometries. The exponent does not depend on the separation or relative dimensions of the oscillator, but the prefactor does. The exponent for the single-rod oscillator is clearly different ($p=2.05$), showing the significance of the presence of two "tines" for acoustic response as opposed to just one. The single-rod oscillator can be thought of emitting acoustic power as a dipole source, whereas the double-rod configuration creates effectively a quadrupole emitter.

\begin{figure}
\begin{center}
\includegraphics[%
  width=0.9\linewidth,
  keepaspectratio]{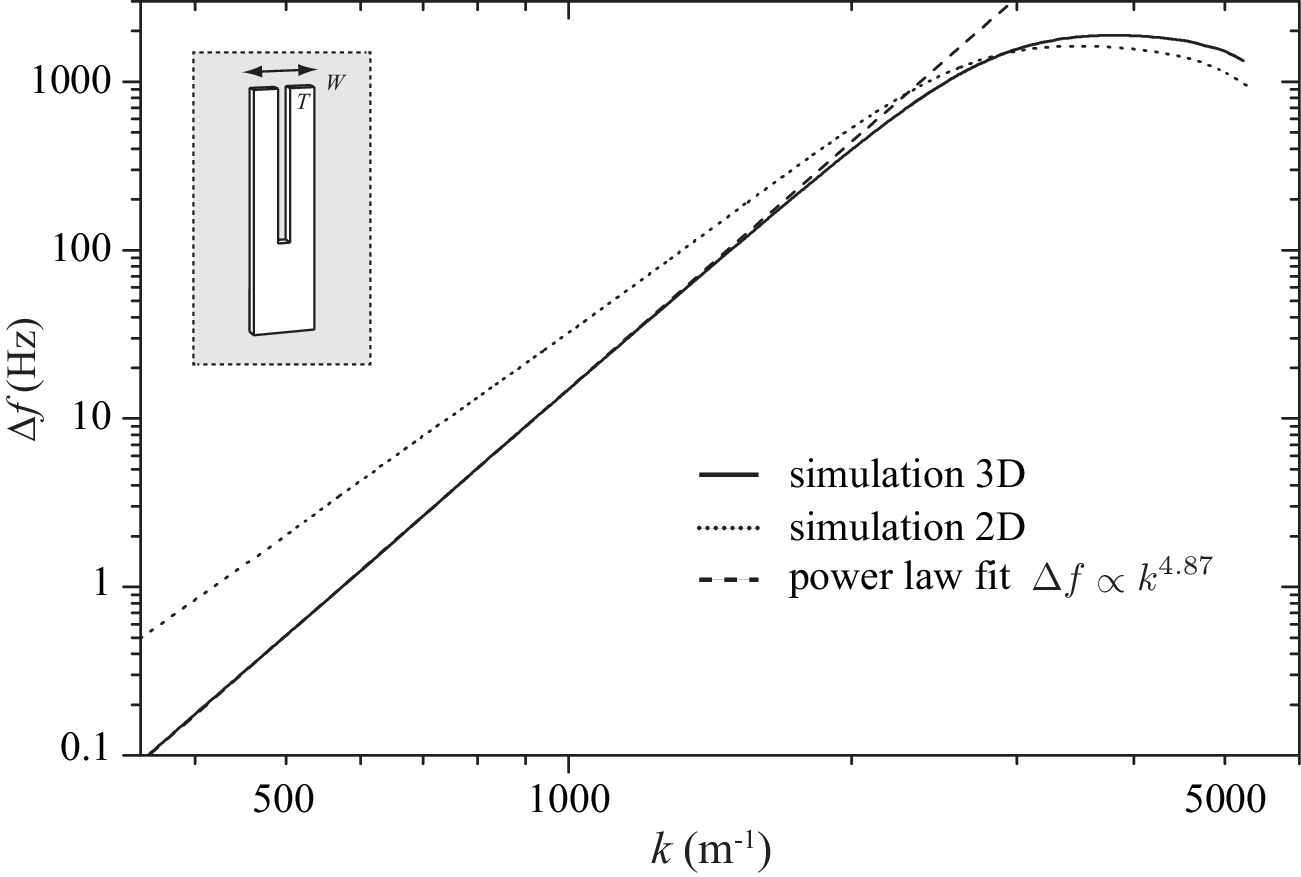}
\end{center}
\caption{Width of the resonance $\Delta f$ as a function of $k$ for a quartz tuning fork in infinite medium. The dimensions are as in Fig.~\ref{Fig:NC38}. Power law fit to the low-$k$ data is plotted as a dashed line. The result for 2D simulations from Fig.~\ref{Fig:AcousticPower2D} corresponding to the tuning fork dimensions is shown for reference (dotted curve). The vacuum width used in the calculation (20~Hz) has been subtracted from $\Delta f$.}
\label{Fig:AcousticPower}
\end{figure}

Resonance width for the 3D tuning fork geometry in infinite medium is represented in Fig.~\ref{Fig:AcousticPower}.
As in the 2D case, which is also reproduced on the figure, a power law fits the simulated data well at low values of $k$. The obtained exponent is significantly larger, $p=4.87$. Such a large exponent is typical for acoustic emission from finite objects. Beyond about $k \approx 1500$~m$^{-1}$, the width starts to deviate from the power law, leveling off and eventually beginning to decrease. Schmoranzer \textit{et al.} have reported on experiments with various quartz tuning forks with different resonant frequencies in superfluid helium, where they found a corresponding exponent $p=5.6$ \cite{David}. Their observations do not rule out an exponent $p \lesssim 5$, however. The authors also constructed several semianalytical models for the acoustic emission from a quartz tuning fork. By modeling the emission as 3D and 2D quadrupole sources and by two infinite cylinders, they obtained $p=6$~(3D), $p=5$~(2D), and $p=5$ (two cylinders). These models are still somewhat crude approximations for the quartz tuning fork and the numerical results presented here are expected to give a more realistic value for the exponent. Bradley \textit{et al.} noted that the 3D model of Schmoranzer \textit{et al.} actually predicts $p=5.5$ \cite{PhysRevB.85.014501}. We should note that our simulated values of $\Delta f$ are somewhat larger at a given wave vector than the experimental results of Refs.~\cite{David} and \cite{PhysRevB.85.014501}. Part of the acoustic energy is reflected back to the oscillator from the walls of the experimental chamber, although they may be at some distance, whereby the excess width decreases, possibly explaining the difference pointed out. Another possibility is that the shapes and sizes of the tuning forks used in the experiments are such that the acoustic emission curves are shifted downward. In our simulations the tuning fork is placed in an environment, where no energy emitted by the oscillator is ever returned. Bradley \textit{et al.} measured the acoustic damping of quartz tuning forks and found $p=5.33$ and $p=5.26$ in $^4$He at $T=4.2$~K and $T=1.55$~K, respectively \cite{PhysRevB.85.014501}. In their experiment only the tine length was varied with the other dimensions kept constant. Our 3D model is more consistent with the experimental data than our 2D simulations.

\subsection{Acoustic Resonances} \label{Sec:AcousticResonances}

\begin{figure}
\begin{center}
\includegraphics[%
  width=0.9\linewidth,
  keepaspectratio]{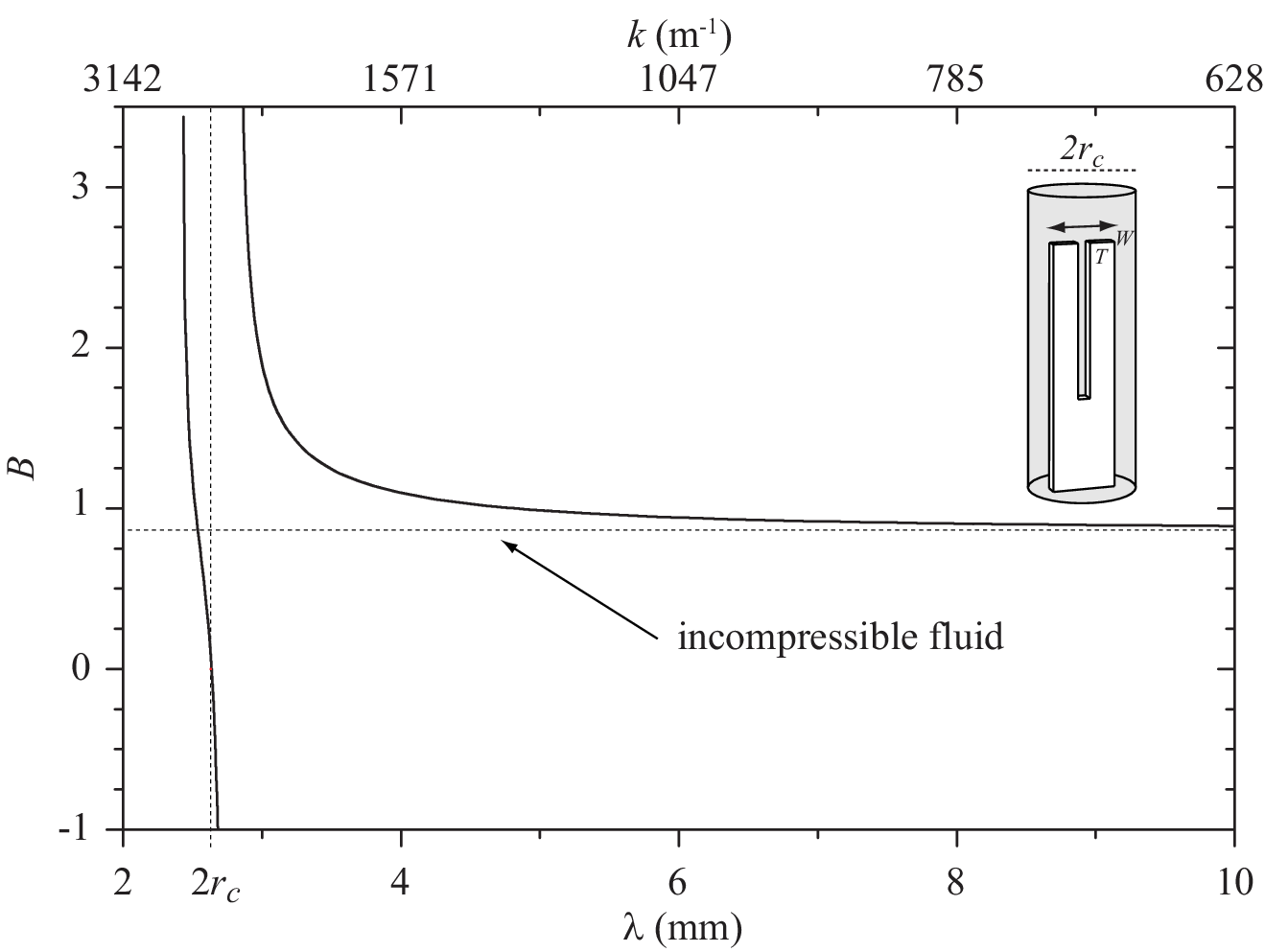}
\end{center}
\caption{Geometrical factor $B$ for a quartz tuning fork inside a perfectly reflecting cylindrical container as a function of wavelength calculated over an acoustic resonance in the cavity. The corresponding wave vector values are indicated on top of the figure. Wavelength $2 r_c = 2.63$~mm (container diameter) is indicated by the vertical dashed line. The tuning fork dimensions are as in Fig.~\ref{Fig:NC38}. The horizontal dashed line is the limit of incompressible fluid, i.e.~$\lambda \rightarrow \infty$. When the resonance is approached from the large wavelength side, $B$ increases steeply and reappears from the negative side. Another resonance is reached soon, causing $B$ to grow rapidly again.}
\label{Fig:BOverResonance}
\end{figure}

When boundaries are added to an acoustic system, new phenomena arise. The mechanical oscillator can couple strongly to acoustic resonances in the fluid cavity. Such resonances significantly affect the behavior of the oscillator. In Fig.~\ref{Fig:BOverResonance} the geometrical factor of a quartz tuning fork positioned symmetrically inside a perfectly reflecting cylindrical cavity is presented, as calculated to larger values of $k$ (smaller wavelengths) than earlier (Fig.~\ref{Fig:BCompareExptCalc}). Now acoustic resonances in the cavity are excited by the tuning fork.
In Fig.~\ref{Fig:BOverResonance}, the data have been plotted as a function of wavelength instead of wave vector to ease comparison with experimental data of Gritsenko \textit{et al.} \cite{Gritsenko}. Applying the acoustic impedance of bulk steel for the walls would not change the results noticeably. We see that the first acoustic resonance appears close to $\lambda \approx 2r_c$, where $r_c$ is the inner radius of the cylindrical container, as one might expect. Close to the resonance, $B$ grows fast and reappears from below, actually achieving negative values. A negative factor means that the resonant frequency of the tuning fork is larger in the fluid than in vacuum. This is consistent with the experiment by Gritsenko \textit{et al.} \cite{Gritsenko}. When the system is close to a fluid resonance, it responds with two distinct resonant frequencies and the behavior of $B$, as seen in Fig.~\ref{Fig:BOverResonance}, is due to the cross-over of the two modes. As $B$ is recovering from the negative side, the system simulated in Fig.~\ref{Fig:BOverResonance} reaches yet another resonance immediately, and $B$ grows again. The "density of modes" increases as $\lambda$ decreases and the behavior of $B$ becomes quite erratic. Since the fork response becomes almost "chaotic" once approaching the emerging plethora of acoustic resonances, it is clear that using the tuning fork for many purposes can then become near impossible. This kind of behavior has been observed in helium mixtures, where so called second sound modes can couple strongly to the quartz tuning forks \cite{Salmela1,Salmela2,SecondSoundModes}. Depending on the application, considering the possible acoustic resonances is worthwhile. If one wants to avoid them as much as possible, removing the container all together may work as a remedy, as long as the tuning fork does not couple too strongly to acoustic modes in the cell geometry. As the fork tine duo can support acoustic resonances in between the two tines as well, the problem is not completely avoidable for this kind of resonators.

To see how many modes there may exist and how strongly they couple to the tuning fork, it is useful to determine the eigenmodes in the fluid-filled cavity. The eigenmodes (pressure field and frequency) are found by solving Eq.~(\ref{Eq:helmholtz}) directly as an eigenfrequency problem with $k^2$ as the eigenvalue. In order to compare different modes with each other, the eigenvectors (pressure fields) have been normalized to the same RMS-value. The relative coupling strength of a particular mode is found by performing a surface integration over the tine surfaces perpendicular to the motion and weighing the integral by Eq.~(\ref{Eq:EulerBeam}). The obtained couplings for the two tines are subtracted, which reflects the demand that the tines are moving in antiphase.

\begin{figure}
\begin{center}
\includegraphics[%
  width=0.9\linewidth,
  keepaspectratio]{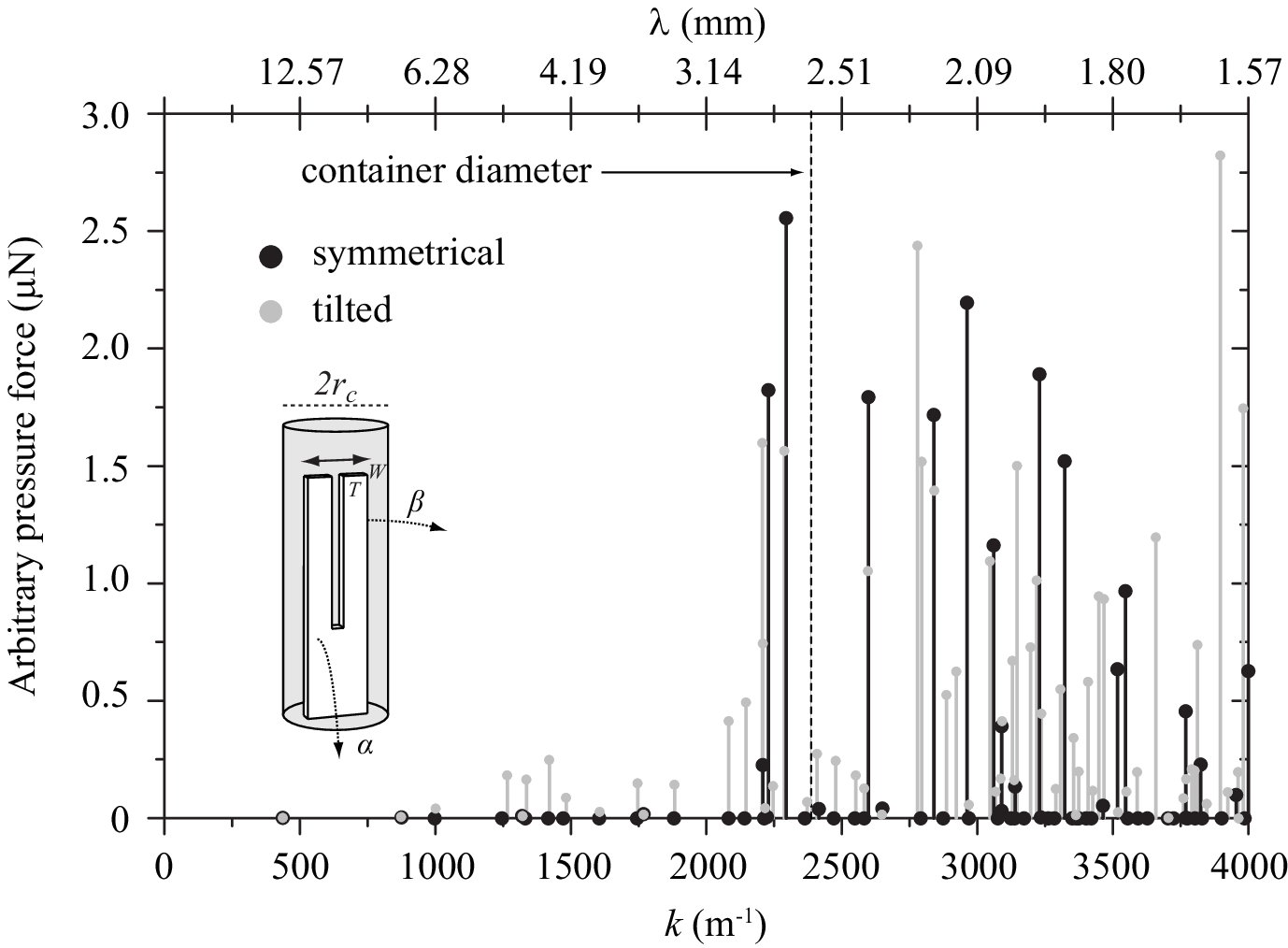}
\end{center}
\caption{Eigenmodes of the fluid cavity when a quartz tuning fork is inside a cylindrical container. Each dot represents an eigenmode at the corresponding wave vector. The vertical axis is the integrated pressure of the fluid on the tines, which represents coupling between the fork oscillation to the corresponding eigenmode of the fluid. Due to only relative norm for the eigenvectors (pressure fields), the scale is arbitrary, whereby just relative strengths between different modes are significant. The horizontal axis shows both the wave vector $k$ (bottom) and the corresponding wavelength $\lambda$ (top). The oscillator dimensions are as in Fig.~\ref{Fig:NC38}. The black dots are for a symmetrically positioned tuning fork and the grey dots are for a fork, which has been tilted by $\alpha = 4^\circ$ in the direction perpendicular to the tine motion and $\beta = 2^\circ$ in the parallel direction. The grey dots are slightly smaller in size and plotted on top of the black dots, so as to display modes, whose couplings do not change. The container diameter ($2r_c = 2.63$~mm) is indicated by a dashed vertical line.}
\label{Fig:ForkEigenvalues}
\end{figure}

Fig.~\ref{Fig:ForkEigenvalues} is a plot of all the acoustic eigenmodes of the cavity in Fig.~\ref{Fig:BOverResonance} up to the wave vector $k = 4000$~m$^{-1}$.
We examine how the modes and the coupling change when the tuning fork is tilted four degrees in the direction perpendicular to the oscillatory motion and two degrees in parallel with it. In the symmetric case the first modes are not excited until $k \geq 2250$~m$^{-1}$, slightly short of the container diameter, as was already indicated in Fig.~\ref{Fig:BOverResonance}. When the fork is tilted, some modes appear already at $k = 1250$~m$^{-1}$. The pressure fields of the lowest-frequency modes ($k<2250$~m$^{-1}$) are such that the nodes are either along the container or in the radial direction perpendicular to the fork oscillation. These are not excited by a symmetrically positioned tuning fork. When the fork is tilted in both directions, the symmetries are broken and some additional modes couple to the oscillator and become visible. The tilt also shifts some of the modes slightly at larger values of $k$.

\begin{figure}
\begin{center}
\includegraphics[%
  width=0.9\linewidth,
  keepaspectratio]{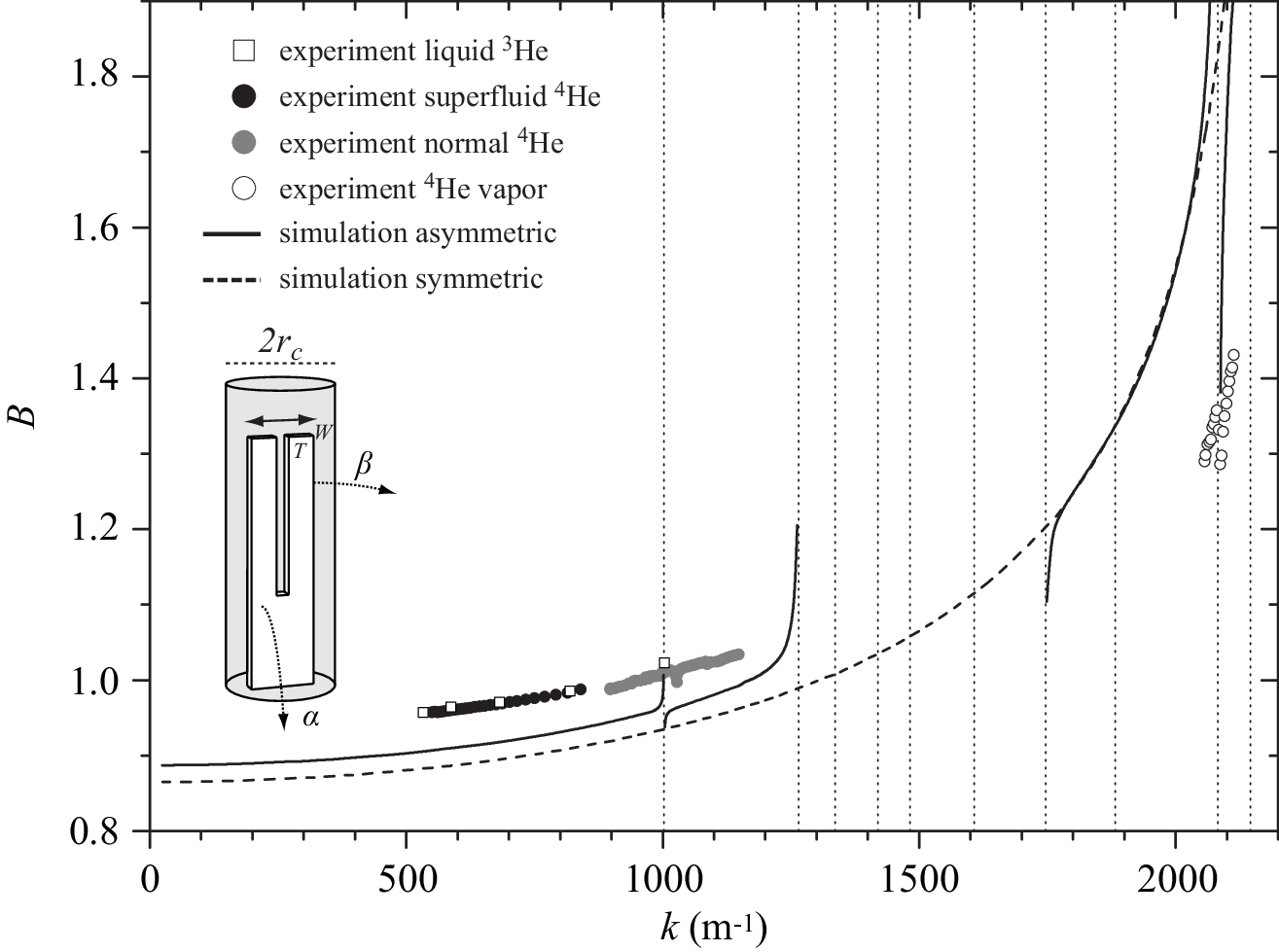}
\end{center}
\caption{Comparison between simulation and experimental data of the geometrical factor $B$ for a quartz tuning fork of the type in Fig.~\ref{Fig:NC38} immersed in helium fluids. The solid line is the simulated value for a fork, which is tilted by $\alpha = 4^\circ$ in the direction perpendicular to the tine motion and $\beta = 2^\circ$ in the parallel direction. The simulation has not been carried through wave vectors from $k=1250$~m$^{-1}$ to 1750~m$^{-1}$ and beyond $k=2100$~m$^{-1}$ due to rapid changes caused by acoustic resonances. Experimental data is missing too, because this region falls in between the accessible wave vectors for liquids and vapor under the examined conditions. The dashed line is the symmetric case from Fig.~\ref{Fig:BCompareExptCalc}. The positions of eigenmodes, whose coupling is strengthened by the tilting are indicated by vertical dotted lines.}
\label{Fig:TiltedComparison}
\end{figure}

We now return to the situation introduced in Fig.~\ref{Fig:BCompareExptCalc} and make comparison of the experimental data with the simulated results for a tilted tuning fork inside a cylindrical cavity using the effective model. The tilt applied is the same as in Fig.~\ref{Fig:ForkEigenvalues}. In this case we must solve the equations of motion for both tines with a common frequency, because the tines can experience different forces from the fluid and can have different amplitudes. Also, we cannot exploit any symmetry conditions and the entire geometry must be meshed in the FEM simulations. This increases the size of elements because of limitations in computing power. Some loss of acoustic energy results, even though the system should be completely lossless. This was observed as an increase in the resonance width, although it should remain at the vacuum value. This did not affect the simulated value of $B$ noticeably, though. The result is represented in Fig.~\ref{Fig:TiltedComparison}.
The geometrical factor has not been calculated between $k=1250$~m$^{-1}$ and 1750~m$^{-1}$, and beyond 2100~m$^{-1}$, due to the number of cavity resonances, since the computation becomes tedious near such resonances. The system behaves in an unstable manner and the set frequency in the "single frequency method" must be very close to the true resonant frequency. The weaker the coupling, the closer one must be to the eigenmode for it to have any significant effect. When the system is close to a cavity resonance, $B$ behaves divergently, as can be seen in this and previous figures.

We note that the cavity resonance at $k = 1882$~m$^{-1}$ does not seem to influence the oscillator too much, even though the coupling strength given in Fig.~\ref{Fig:ForkEigenvalues} suggests stronger coupling than for the $k=1002$~m$^{-1}$ mode, for example. By running the simulation with very small wave vector steps in the vicinity of that mode, its coupling to the oscillator can be recognized. The reason for the seemingly large coupling may be that the normalization used for the eigenvectors is not perfect and for some pressure fields the procedure of calculating the coupling overestimates the actual situation. Similarly the coupling for some modes may be underestimated. If a mode is such that the pressure field tends to move the tines to the same direction, which is the case with the $k=1882$~m$^{-1}$ mode, the computed coupling term becomes a difference between two rather large numbers. In this case mesh quality, for example, begins to have greater importance. Our computed coupling term takes into account only how the fluid modes affect the tines, but may not take correctly into consideration how the mode is excited by the tuning fork in the first place. Since the tines must oscillate in antiphase, a mode in which the tines should move in the same direction, cannot be excited very efficiently. If the fluid forces of such a mode acting on the tines are different in magnitude, the oscillator has two resonance modes nearby, which differ in phase by 180 degrees.

The small bump in the experimental data at $k \approx 1025$~m$^{-1}$ could be the first additional mode to be excited by the tilted fork, which in our simulated configuration appears at $k = 1002$~m$^{-1}$. We find other modes around $k = 2100$~m$^{-1}$, where the $^4$He vapor data exist, which can explain the low values of the experimental $B$ compared to the symmetrical simulation results. The closer proximity of the cylinder walls due to tilting of the oscillator is apparently the reason for the increase in $B$ at small wave vectors compared to the symmetrical case. At larger $k$ the $B$-factor seems to be closer to that of a symmetrical positioning, unless a cavity resonance is nearby. One can imagine, that by tuning the modeled system by varying the fork position and tilt, simulated data could be made to fit the experiment even better.

Since no other plausible effects have come up and the one examined NC38 quartz tuning fork had a clearly observable tilt, we conclude that the discrepancy between the symmetrical simulation and experiment is mainly due to the tilted fork leaning towards the wall and exciting additional modes inside the cavity. It would be interesting to obtain experimental data in the wave vector range between $k=1150$~m$^{-1}$ and $k=2050$~m$^{-1}$. This could be covered, for example, with pure $^3$He between 1~K and 3~K, where the speed of sound changes between 180~m/s and 100~m/s. At these temperatures, viscosity of $^3$He is not excessive.

\section{Conclusions}

We have performed numerical simulations to study the characteristics of quartz tuning fork resonators in acoustic medium. We have verified that the "effective model" described in Sec.~\ref{Sec:Fork} is accurate to within a few percent compared to a more elaborate "full model". The effective model offers computationally cheaper way to simulate the tuning fork in fluid. It can be extended to other vibrational modes of the fork besides the lowest-frequency mode by determining the displacement as a function of height for the mode in question, similar to Eq.~(\ref{Eq:EulerBeam}). The higher-frequency modes require some modification to the harmonic oscillator approximation \cite{PhysRevB.85.014501}. A small discrepancy between the two models remains and cannot be explained with absolute confidence at this moment. A reasonable agreement with an available set of experimental data has also been achieved. Given the uncertainties in the exact experimental geometry, the agreement could be considered quite good. More experimental data with different oscillators would be of interest. Acoustic emission has been shown to become an important effect with quartz tuning forks in helium, when the wave vector increases (small $\lambda$), since the emitted acoustic power is proportional to $\sim k^5$.

Our results can help choosing the appropriate quartz tuning forks for various experiments and deciding how to mount them. In many applications, an oscillator with a constant geometrical parameter would be desirable. Our 2D simulations indicate that the tines should be as "narrow" as possible to achieve this. That is, the ratio $T/W$ should be large. However, this increases the emitted acoustic power at lower wave vectors (decreases at larger $k$). The exponent in the acoustic-emission law remains unchanged. In some applications a large surface area in the direction of motion relative to the oscillator mass is advantageous for higher sensitivity. In this case maximizing the $T/W$ ratio may not be optimal.

We have also demonstrated that acoustic resonances in the container cavity play a significant role, even at wavelengths much larger than the container diameter. To avoid low-$k$ acoustic resonances in the cavity, the quartz tuning fork should be installed symmetrically in the container. We remind that in addition to first sound, second sound in superfluids can also couple to a mechanical resonator, especially in helium mixtures. Since the speed of second sound is an order of magnitude smaller than that of first sound, these resonances appear at lower oscillator frequencies than the first sound modes. Second sound effects have not been examined in this work.

Our results strictly apply to inviscid fluids only. Viscosity introduces another geometrical parameter, which can be included in Eq.~(\ref{Eq:Bdefinition}). In the present treatment, viscosity could be accounted for as a linear approximation by adding complex-valued terms into the wave equation, Eq.~(\ref{Eq:helmholtz}). In some cases, such as those presented in Refs.~\cite{springerlink:10.1007/s10909-007-9583-7} and \cite{PhysRevB.78.064509}, experimental data can be extrapolated to zero viscosity, and Eqs.~(\ref{Eq:Bdefinition}) and (\ref{Eq:helmholtz}) are valid as such.

\begin{acknowledgements}
We thank F.\ B.\ Rasmussen for useful comments. This work has been supported in part by the EU 7th Framework Programme (FP7/2007-2013, Grant No.\ 228464 Microkelvin) and by the Academy of Finland through its LTQ CoE grant (project no.\ 250280). We also thank the National Doctoral Programme in Materials Physics for financial support.
\end{acknowledgements}

\bibliographystyle{apsrev4-1}
\bibliography{refs}   

\end{document}